\documentclass[11pt,a4paper]{scrartcl}

\usepackage[utf8]{inputenc}
\usepackage[T1]{fontenc}
\usepackage[english]{babel}
\usepackage[autostyle]{csquotes}

\usepackage{lmodern}
\usepackage[pdftex]{graphicx}
\usepackage{pdfpages}

\usepackage[noblocks]{authblk}
\usepackage{orcidlink}
\usepackage[nointegrals]{wasysym}
\usepackage{fontawesome5}

\usepackage[all]{nowidow}
\usepackage{multicol}
\usepackage{array}
\usepackage{amsmath} % needed for table legend
\usepackage{booktabs} % better looking tables
\usepackage{threeparttable} % needed for better table layouts
\usepackage{hologo} % Provides logos for, e.g., BibTeX

\usepackage{float}
\usepackage{subcaption}
\usepackage{caption}

\usepackage[margin=1.5cm, includefoot]{geometry}
\usepackage{fancyhdr}
\usepackage{enumitem}
\setlist[itemize]{itemsep=0.1em, parsep=0em, topsep=0.2em}
\setlist[enumerate]{itemsep=0.1em, parsep=0em, topsep=0.2em}
\setlist[description]{itemsep=0.3em, parsep=0em, topsep=0.2em, align=parright, leftmargin=!}
\SetLabelAlign{parright}{\strut\smash{\parbox[t]{\labelwidth}{\raggedleft#1}}}

\usepackage{tcolorbox} % used for community feedback hint
\usepackage[colorinlistoftodos,prependcaption]{todonotes} %% REMOVE BEFORE FINAL RENDERING

\usepackage[style=ieee]{biblatex}
\usepackage[toc,section=section]{glossaries}

\usepackage{hyperref}
\hypersetup{
    pdfauthor={Bouquin, D.; Bertuch, O.; Colon-Marrero, E.; Trisovic, A.},
    pdfkeywords={Research software; Software publication; Software metadata; Repositories; Metadata extraction; Metadata harvesting; Metadata formats; FAIR; FAIR4RS},
    pdflang={EN},
    bookmarksopen=true,
    breaklinks=true,
    colorlinks=true,
    linkcolor=blue,
    citecolor=blue,
    filecolor=blue,
    urlcolor=blue
}

\usepackage{caption}
\usepackage{bookmark}
\makeatletter
    \pretocmd\endtable{%
      \bookmark[
        rellevel=1,
        keeplevel,
        dest=\@currentHref,
      ]{Table \thetable: \@currentlabelname}%
    }{}{\errmessage{Patching \noexpand\endtable failed}}
    \pretocmd\endfigure{%
      \bookmark[
        rellevel=1,
        keeplevel,
        dest=\@currentHref,
      ]{Figure \thefigure: \@currentlabelname}%
    }{}{\errmessage{Patching \noexpand\endfigure failed}}
    
    \AtBeginDocument{
        \lfoot{\enquote{\@title}}
        
        \hypersetup{
            pdftitle={\@title},
            pdfsubject={\@subtitle: ...}
        }
    }
\makeatother

\usepackage{crossreftools} % used to better control text of namerefs, load after hyperref to let it create linked refs!

\title{Advancing Software Citation Implementation}
\subtitle{Software Citation Workshop 2022}

\date{\today}
\publishers{
    \vspace{0.5cm}
    \small
    With contributions from 
    
    Stephen Abrams, Daniel Chivvis, Stephan Druskat, \\
    Mark Hahnel, Tom Honeyman, Daniel Katz, \\
    August Muench, Vicky Rampin, and Jurriaan Spaaks
}

\author[1,*]{\orcidlink{0000-0003-2626-3688} Bouquin, Daina}
\author[2,*]{\orcidlink{0000-0003-1991-0533} Trisovic, Ana}
\author[3]{\authorcr \orcidlink{0000-0002-2702-3419} Bertuch, Oliver}
\author[4,5]{\orcidlink{0000-0002-9855-4081} Colón-Marrero, Elena}

\affil[1]{Harvard-Smithsonian Center for Astrophysics, Cambridge, USA}
\affil[2]{Institute for Quantitative Social Science, Harvard University, Cambridge, USA}
\affil[3]{Forschungszentrum Jülich GmbH, Germany}
\affil[4]{Bentley Historical Library, University of Michigan, Ann Arbor, USA}
\affil[5]{Software Preservation Network}
\affil[*]{Corresponding Author}

\makeglossaries
\newglossaryentry{GLAM}{
   name=GLAM,
   text=GLAM,
   description={is an abbreviation of \enquote{\textbf{G}alleries, \textbf{L}ibraries, \textbf{A}rchives, \textbf{M}useums} and defines the \enquote{four types of cultural institutions that share a collaborative relationship stemming from their similar missions} (taken from the \href{https://dictionary.archivists.org/entry/galleries-libraries-archives-and-museums.html}{SAA Dictionary})}
}

\begin{document}
\maketitle
\thispagestyle{empty}

\begin{abstract}\label{abstract}
\addsec{Abstract}
Software is foundationally important to scientific and social progress, however, traditional acknowledgment of the use of others' work has not adapted in step with the rapid development and use of software in research. 

This report outlines a series of collaborative discussions that brought together an international group of stakeholders and experts representing many communities, forms of labor, and expertise. Participants addressed specific challenges about software citation that have so far gone unresolved. The discussions took place in summer 2022 both online and in-person and involved a total of 51 participants. 

The activities described in this paper were intended to identify and prioritize specific software citation problems, develop (potential) interventions, and lay out a series of mutually supporting approaches to address them. The outcomes of this report will be useful for the GLAM (Galleries, Libraries, Archives, Museums) community, repository managers and curators, research software developers, and publishers.
\end{abstract}

\begin{center}
    \vspace*{\fill}
    \footnotesize
    This work is licensed under \href{https://creativecommons.org/licenses/by/4.0/}{\faCreativeCommons\faCreativeCommonsBy\,4.0}.    
\end{center}
\clearpage

\tableofcontents
\begin{center}
    \vspace*{\fill}
    \begin{tcolorbox}[colback=blue!5!white,colframe=blue!75!black,width=\linewidth-2cm]
        \paragraph{Community feedback}\label{community-feedback}
        
        At this stage, we would like to reach out to the community to gain insights into what desirable solutions can 
        look like, what potential pitfalls are, etc.
        
        We kindly ask readers and other interested parties to provide feedback on the concept detailed here via its PubPeer page at
        \url{https://go.fzj.de/peerpub-scw22}.
    \end{tcolorbox}
\end{center}

\clearpage
\section{Introduction}\label{sec:introduction}
Software has been a crucial intellectual contribution to scientific and social progress for decades. However, traditional acknowledgment of the use of others’ work has not adapted in step with the rapid development and use of software in research. In particular, standard machine-actionable citations have been inconsistently applied to software or not at all. While policies regarding software citation are now changing (as tracked online at \href{https://www.chorusaccess.org/resources/software-citation-policies-index}{CHORUS}), practice is slow to follow policy. 

Studies have shown that researchers reference software in their publications using a variety of inconsistent and ambiguous mechanisms. For example, a twenty-year case study analyzing software citation behavior in astronomy found that 109 different aliases (including alternate names, phrases, and publications) were used to identify just nine software packages in papers from a single publisher, and hundreds of papers mentioned the software without giving any recognizable form of attribution to software authors \cite{bouquin_credit_2020}. Similarly, a study examining software citation in a random sample of biology publications \cite{howison_sw_lit_2016} found \enquote{diverse and problematic practices}, emphasizing that software is very often only informally mentioned, even in high-impact factor journals. The same study found that current software citations often fail to accomplish the basic functions of citation. These studies highlight the need for better support for software citation.

An alternative way to cite software is by proxy using (software) papers \cite{chue_hong_ssi_journals_2012, van_noorden_top100_2014}, documentation on websites, and discipline-specific registries \cite{bouquin_credit_2020}. For example, the Journal of Open Source Software (JOSS) was created in part to allow for a paper placeholder that could be cited in place of software, while also improving the software quality with more documentation and reporting \cite{smith_joss_2018}.

Recognizing the need for software citation, a FORCE11 working group developed the Software Citation Principles in 2016 \cite{smith_sw_cit_prin_2016}. FORCE11 working groups are international, interdisciplinary collaborations, involving a variety of diverse stakeholders who contribute to shaping the future of scholarly communication. The FORCE11 Software Citation Principles established core values and motivations for software citations:

\begin{description}[labelwidth=4.5cm, rightmargin=1cm]
    \item[Importance] Software should be considered a legitimate and citable product of research.
    \item[Credit and attribution] Software citations should facilitate giving scholarly credit and normative, legal attribution to all contributors to the software.
    \item[Unique identification] A software citation should include a method for identification that is machine actionable, globally unique, interoperable, and recognized.
    \item[Persistence] Unique identifiers and metadata describing the software and its disposition should persist.
    \item[Accessibility] Software citations should facilitate access to the software itself and to its associated metadata.
    \item[Specificity] Software citations should facilitate identification of, and access to, the specific version of software that was used.
\end{description}

The Software Citation Principles are not on their own a solution to the issues impacting software citation implementation and application. Instead, the Software Citation Principles should be seen as the first step toward a goal. An important next step was taken by a group called the FORCE11 Software Citation Implementation Working Group (SCIWG) \cite{katz_f11_sciwg_2022}, which has a goal to direct relevant people towards a proposed best practice, while also assessing, documenting and addressing predominant challenges, including technical and cultural challenges, that prevent the software citation principles from becoming a reality \cite{katz_sw_impl_chall_2019}. 

A related effort, aiming to improve research software by making it Findable, Accessible, Interoperable, and Reusable (FAIR) \cite{wilkinson_fair_2016} online, was proposed by a Research Data Alliance working group called \enquote{\href{https://www.rd-alliance.org/groups/fair-research-software-fair4rs-wg}{FAIR for Research Software}} (FAIR4RS) in 2020. The group organized a number of meetings and discussions, leading to comprehensive reports on FAIR for research software \cite{chue_hong_fair4rs_2022,katz_fresh_look_2021}. 

In order to advance software citation and facilitate discussion, the John G. Wolbach Library sought and obtained funding from the Institute for Museum and Library Studies to hold a set of online and in-person events to bring together stakeholders representing the various forms of labor and expertise necessary to make progress on implementing the Software Citation Principles. In addition to computer scientists and researchers, the event aims to gather \gls{GLAM} professions, who are strategically well-situated to address issues impacting how and if software is disseminated, discovered, and the extent to which metadata about software is standardized (e.g., adoption of schemas like CodeMeta \cite{jones_codemeta_2017}, Citation File Format (CFF) \cite{druskat_cff_2021}; software ontologies \cite{malone_softonto_2014, garjo_softdesconto_2021, wilder_doap_2021, wuersch_seon_2012}). Such a diverse group of attendees has the potential to tackle cultural issues impacting software citation,  ensure that platforms (e.g., archives, data repositories, registries) address their community’s software dissemination needs, bridge divides between potential collaborators, and develop new tools that support authors.
\section{Methodology: Software Citation Implementation Workshops}\label{sec:methodology}

The events were established as Software Citation Workshop 2022 and \href{https://library.cfa.harvard.edu/software-citation-workshop-2022}{advertised on the web}.
Its online and in-person events were designed to encourage discussion and collect participants’ input.
The online activities or “online focus groups”, were intended to inform the direction of the in-person workshop and to ensure that a large group of stakeholders were able to contribute their thoughts.\footnote{Restrictions implemented in response to the COVID-19 pandemic required organizers to keep the in-person meeting small (no more than 15 attendees)}
They also allowed organizers to bring together diverse perspectives from people who could not travel, or chose not to, while accommodating Earth-spanning time zones.

\subsection{Online Focus Groups}\label{subsec:online-focus-groups}

The online focus groups revolved around three goals, which were:

\begin{itemize}
    \item to understand how people view the problem of software citation from various perspectives, 
    \item to identify barriers to implementing software citation metadata standards, and 
    \item to learn how people prioritize problems in this context.
\end{itemize}

In order to meet these goals, the organizers worked with a professional facilitator\footnote{Dialogic Consulting provided professional facilitation services: \url{https://www.dialogicconsulting.com}} to develop discussion prompts and follow-up questions that would guide the participants through productive conversations on this complicated topic.
A full “run of show” was established (\href{https://osf.io/aby75}{Appendix 1}) and a graphic note-taker was commissioned to record the sessions.
The graphic note-taker’s work was used to visually represent the discussion and keep participants on topic (\href{https://osf.io/tu7j8}{Appendix 2}).
Online focus groups were held over the course of three days in August 2022, gathering 45 people in total, who participated across three focus group sessions (\href{https://osf.io/hxy5k}{Appendix 3}).

The following conclusions were synthesized after the online focus groups were completed: 

{
\renewcommand{\theenumi}{\textsl{\arabic{enumi}}}
\begin{enumerate}%[labelsep = 0.2em]
    \item \textsl{Most preservation platforms were designed for data and papers, but not software.} \\
    Shoehorning software into systems and procedures designed for different digital objects makes enabling software citation difficult. Workflows, staff practices, metadata fields, and tools were all developed without considering software.
    
    \item \textsl{Awareness of software citation metadata standards is low.} \\
    Adoption of these standards also remains limited\footnote{For CFF, see \url{https://github.com/sdruskat/cfftracker} for usage on Github}, with unclear influence of (non-)awareness.
    
    \item \textsl{Multiple standards with different strengths can cause confusion.} \\
    The community generally agreed that any citation file is good, but two standards can make conversations and adoption less straightforward.
    
    \item \textsl{There are very few immediate incentives.} \\
    Encouraging software authors to generate and curate software citation metadata is hard. Long-term benefits are less likely to motivate action than near-term benefits.
    
    \item \textsl{Automatically generated software citation metadata is very messy.} \\
    We need tools like HERMES\footnote{\cite{hermes_2022}, see also \url{https://software-metadata.pub}} that make citations easier to generate and the curators to maintain the metadata from one software version to the next.
    
    \item \textsl{Tooling to support software citation can be buggy.} \\
    If tools don’t work on the first try, people are unlikely to try using them again.
    
    \item \textsl{There is already software in archives published alongside datasets with missing metadata.} \\
    Enabling software citation when the software and the metadata are no longer maintained is a challenge.
    
    \item \textsl{It is not clear when to create separate records for software and its related digital objects.} \\
    When is it better to create multiple identifiers? How do you get people to create multiple archival deposits?  Is it worth it to create separate records for even the tiniest research software?
    
    \item \textsl{Who does the work?} \\
    Metadata curation needs to continue as long as the software is actively developed, but whose responsibility it should be to do that maintenance is unclear.
\end{enumerate}
}

The summary from each focus group session was shared with focus group participants and in-person workshop attendees. These themes were used to shape conversations throughout the in-person event.

\subsection{In-Person Workshop}\label{subsec:in-person-workshop}
Informed by the focus group insights, the in-person workshop had the following objectives: 

\begin{itemize}
    \item Brainstorm interventions to tackle software citation problems
    \item Prioritize problems and lay out a series of approaches to address relevant problems
    \item Identify the resources necessary to implement solutions
    \item Synthesize a plan of action and other deliverables to be shared widely. 
\end{itemize}

In August 2022, thirteen people from five countries gathered at the Center for Astrophysics in Cambridge, MA to meet the workshop objectives.
In order to accomplish the goals outlined above, attendees participated in a series of structured brainstorming sessions and follow-on activities in small groups.
The idea was to optimize creative problem-solving and practical application of ideas to move the needle on software citation.

\subsubsection{Areas of Focus}\label{subsubsec:areas-of-focus}
At the start of the session, all participants were asked to introduce themselves and to engage in an open discussion about the major themes identified during the online focus groups. During this time, participants refined their ideas and zeroed in on the challenges that would be centered throughout the workshop. In the end, three groups were formed, with each focusing on a specific area of interest:

\paragraph{Group 1 - A vision of software citation}\label{par:focus:group1}

It became clear during our initial conversation that there is no cohesive vision for the future of software citation. Without a unified vision, it is challenging to advocate for or enact systemic changes across disciplines and stakeholder groups to address any of the issues that surfaced during the focus groups. Because software citation is a highly interdisciplinary area of interest, this lack of direction acts as a hindrance to all forward progress. For example, there are mixed opinions about the role and sustainability of software journals in the context of software citation. A broad consensus has been reached on data publication practices (typically by sharing data in data repositories and registries) but a similar consensus has not yet been reached by the software citation community.

\paragraph{Group 2 - Actions that can be taken without additional resources}\label{par:focus:group2}

There are many social and technological challenges impacting software citation implementation, and many ideas for addressing them require additional resources (e.g., staff time, finances, institutional support). Despite this, the group decided to spend time developing ideas and plans for actions that could have a positive impact with the resources that are available right now.

\clearpage
\paragraph{Group 3 - Interdisciplinary resources}\label{par:focus:group3}

Through conversations between \gls{GLAM} professionals and others within the scholarly communication ecosystem, it became clear that people within different fields use different terminology and have different ways of thinking about the issues outlined in the major themes from the focus groups. For this reason, the third in-person group decided to discuss resources that could be used to make software citation issues more relevant and actionable within \gls{GLAM} settings.

\paragraph{Unattended topics}
The following were other important topics that participants thought warranted focus, but in the end there was more interest in pursuing the three areas outlined above:

\begin{itemize}
    \item \textsl{Determining the scale of the problems impacting software citation implementation} \\
          For example, establishing a way to understand where we’re starting and to quantify progress.
    \item \textsl{Developing an approach to addressing software citation in the context of mixed data types.}
          \begin{description} [labelwidth=1cm, topsep=0pt]
               \item[Note:] This issue is discussed further in the Group 3 Outcomes section
               \ref{subsec:outcome:group3} below. \\
                Conversation on this topic made it clear that additional resources are needed to support communication between \gls{GLAM} professionals and other stakeholders working on software citation before much progress could be made on mixed data types.
    \end{description}
\end{itemize}

\subsubsection{Methodology: Structured Brainstorming}\label{subsubsec:brainstorming}
After settling on the topics described above, the three groups undertook a series of structured brainstorming sessions, each building upon the last, to determine a path forward.

\paragraph{Session 1: \enquote{Disney Method}}\label{par:method:session1-disney}

\begin{figure}[H]
    \begin{subfigure}[c]{0.5\textwidth}
        \includegraphics[width=0.95\linewidth]{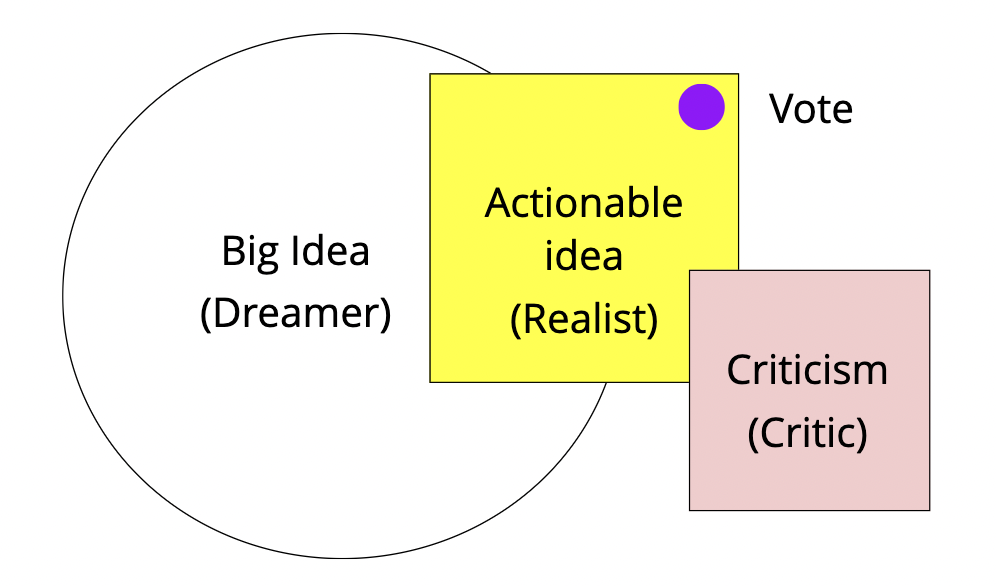}
    \end{subfigure}
    \begin{subfigure}[c]{0.5\textwidth}
        \includegraphics[width=0.95\linewidth]{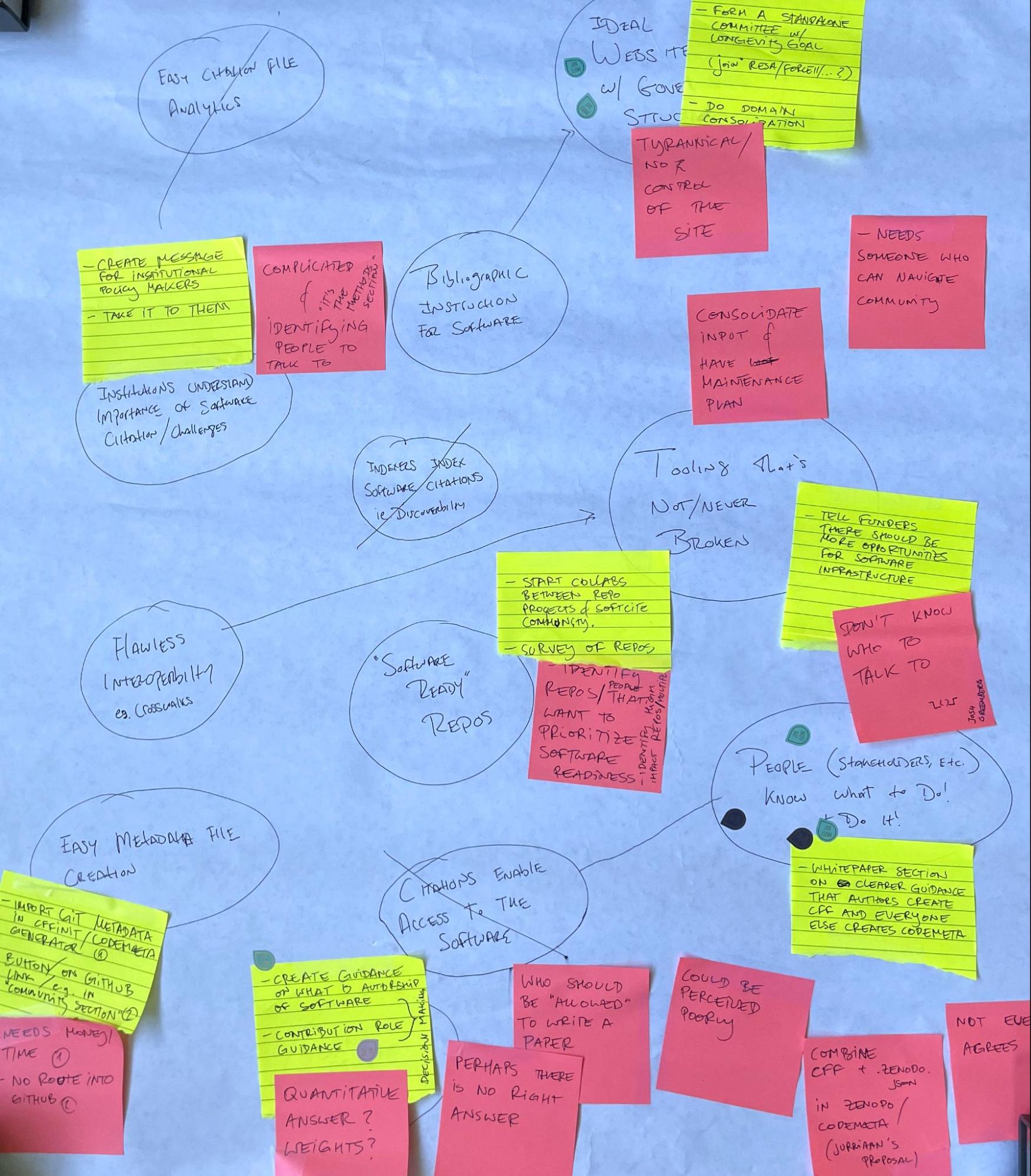}
    \end{subfigure}
    
    \caption{
    Brainstorming activities from \enquote{Disney Method} session (see \ref{par:method:session1-disney}) were done on paper (right) as templated (left) during the session's introduction.
    }
    \label{fig:disney-method}
\end{figure}

\href{https://www.designorate.com/disneys-creative-strategy/}{The \enquote{Disney} strategy} (see figure \ref{fig:disney-method}) is a multi-step creativity strategy in which groups use a series of specific thinking styles to analyze a problem and then generate and refine actionable ideas. The thinking styles are broken into three \enquote{personas} that participants are asked to adopt throughout the exercise:

\begin{description}[labelwidth=3cm, rightmargin=1cm]
    \item[The Dreamer] When adopting this persona, attendees focus on generating ideas without thinking about constraints.
    \item[The Realist] Participants adopt this persona to refine their ideas to make them more discrete and actionable.
    \item[The Critic] The last persona is adopted to identify flaws in the ideas that have been generated with the goal being to avoid failure (constructive criticism).
\end{description}

After each group goes through the process of adopting all three personas for a limited period of time and recording their thoughts, everyone votes on their top two ideas to take to the next session of brainstorming.

\begin{figure}[H]
    \begin{subfigure}[c]{0.5\textwidth}
        \includegraphics[width=0.95\linewidth]{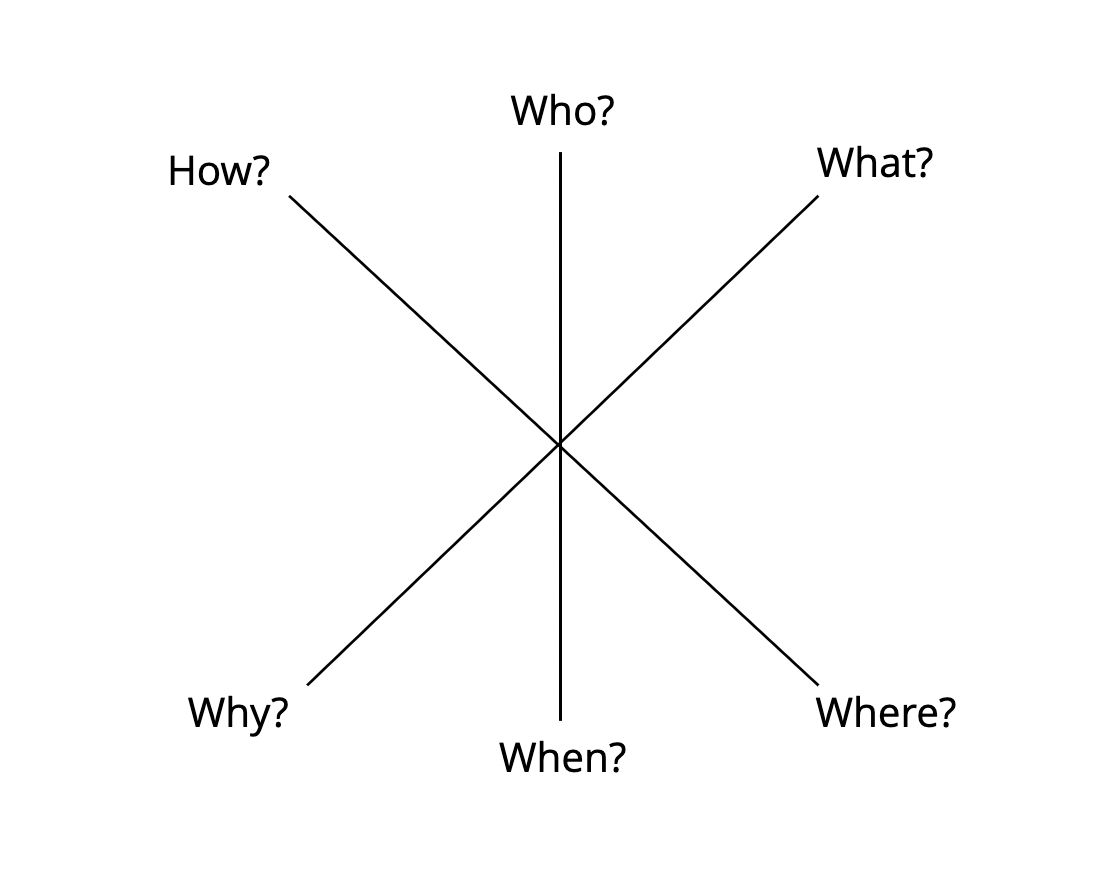}
    \end{subfigure}
    \begin{subfigure}[c]{0.5\textwidth}
        \includegraphics[width=0.95\linewidth]{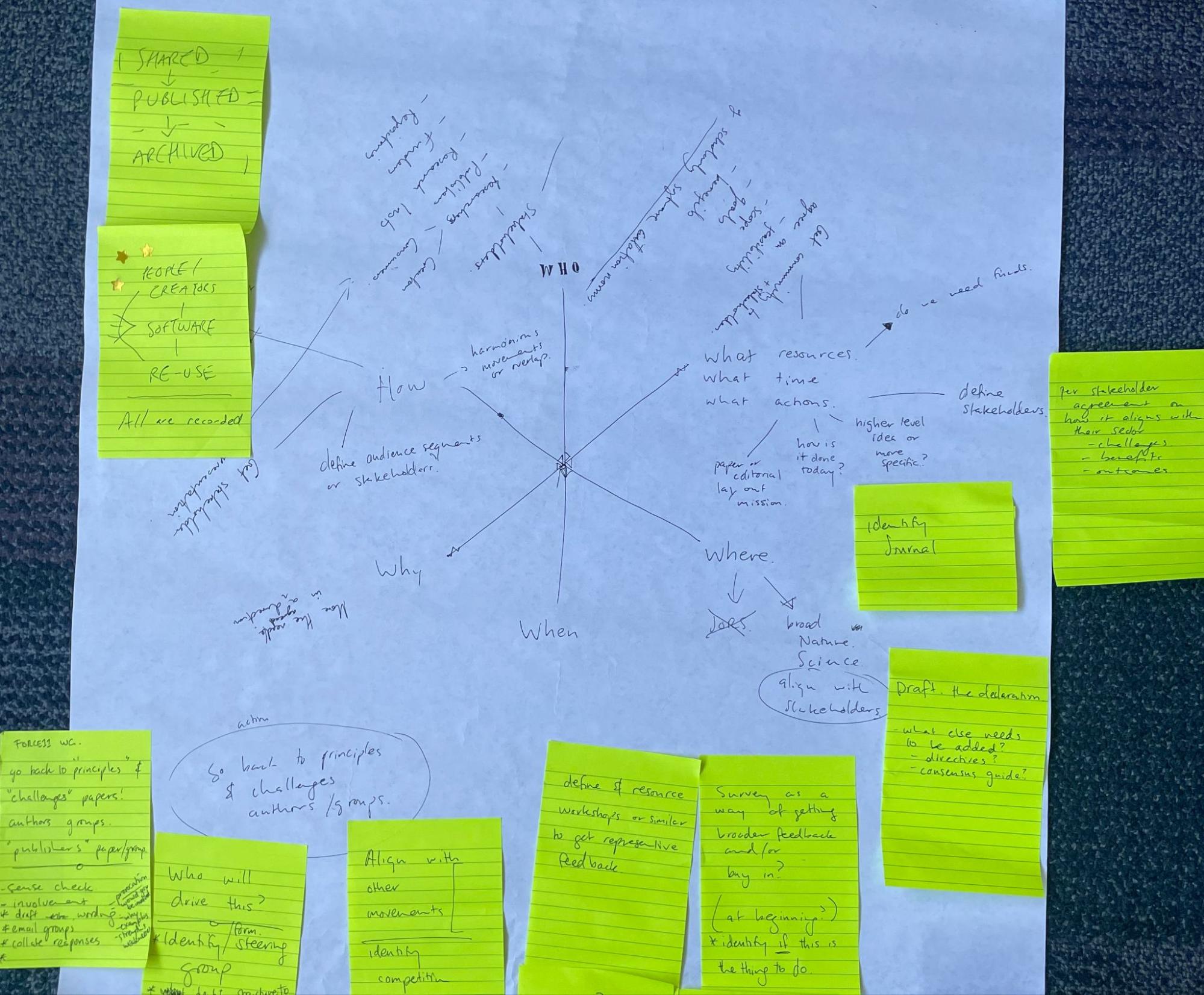}
    \end{subfigure}
    
    \caption{
    Participants spent time determining what questions would need to be answered in order to execute their ideas before working toward answers. The \enquote{starbursting} exercise (left, see \ref{par:method:session2-starburst}) was executed on paper during the session (right).
    }
    \label{fig:starbursting}
\end{figure}

\paragraph{Session 2: Starbursting and Gap Filling}\label{par:method:session2-starburst}
Building on the voted outcomes of the Disney method in session 1, groups participated in a \enquote{starbursting} exercise. \href{https://lucidspark.com/blog/how-to-use-starbursting-for-brainstorming}{Starbursting} (see figure \ref{fig:starbursting}) is used to generate questions rather than answers and focus on identifying the \enquote{who, what, where, when, why, and how} of executing any given idea.

Once these questions were generated, participants spent time answering the questions. This series of exercises allowed people to start identifying actionable paths forward for their ideas and determining the resources and barriers associated with them.

\begin{figure}[H]
    \begin{subfigure}[c]{0.33\textwidth}
        \includegraphics[width=0.95\linewidth]{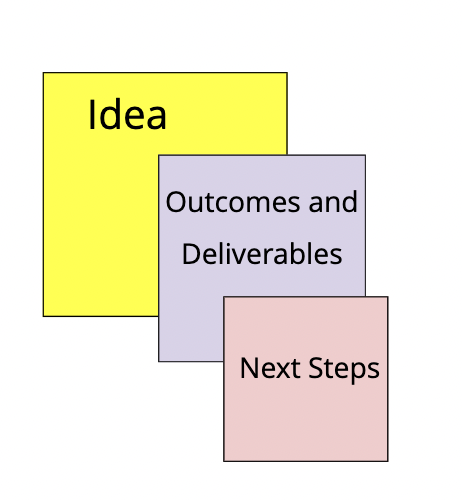}
    \end{subfigure}
    \begin{subfigure}[c]{0.66\textwidth}
        \includegraphics[width=0.95\linewidth]{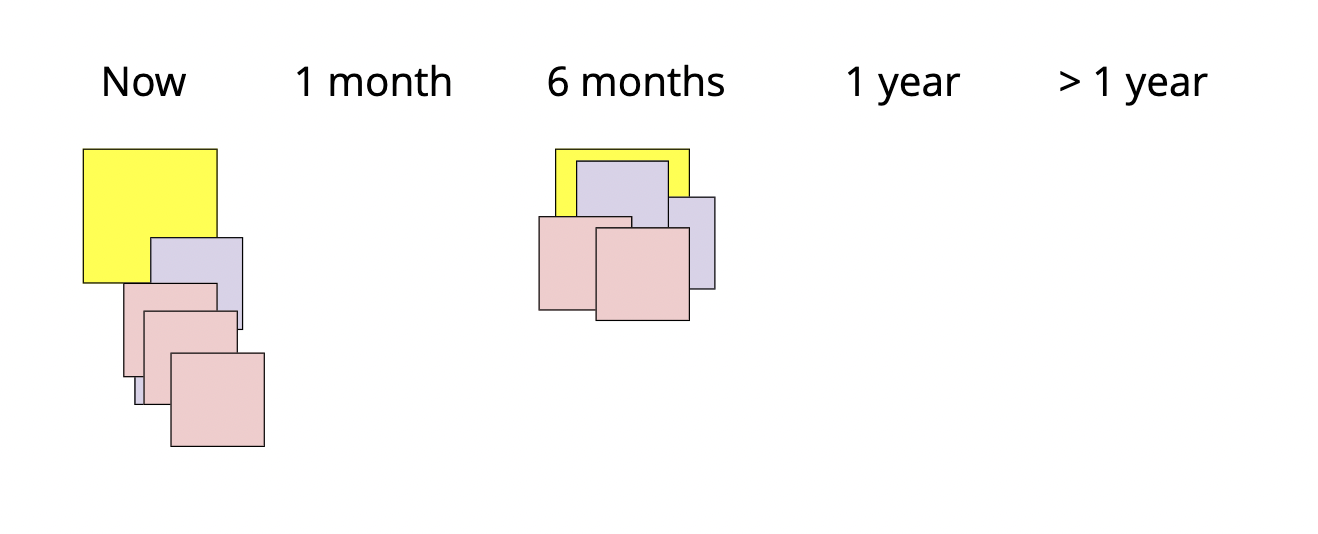}
    \end{subfigure}
    
    \caption{
    Roadmapping process (see \ref{par:method:session3-roadmap}) undertaken by each group.
    Idea times with potential outcomes and next steps (left) where places on a timeline (right).
    }
    \label{fig:roadmapping}
\end{figure}

\paragraph{Session 3: Road Mapping}\label{par:method:session3-roadmap}
The final structured brainstorming activity was \href{https://www.officetimeline.com/roadmaps}{\enquote{road mapping}} (see figure \ref{fig:roadmapping}). At this stage, all groups were asked to lay out their ideas on a timeline and then write outcomes, deliverables, and dependencies that would be associated with each idea.

These steps were iterative and required groups to adjust their timelines depending on the deliverables and outcomes they envisioned. The final stage of this exercise required groups to identify the next steps to achieving progress. Each group’s roadmap was informed by the prior starbursting and gap-filling activities.
\clearpage
\section{Outcomes and Discussion}
Photos of paper-based outputs following the methods in sec. \ref{subsubsec:brainstorming} have been \href{https://osf.io/ad65z}{deposited to OSF}.

\subsection{Group 1: A vision for software citation}\label{subsec:outcome:group1}
Today, there are many stakeholders interested in various aspects of software citation:

\begin{multicols}{2}
    \begin{itemize}
        \item Research software creators and users
        \item Publishers
        \item \gls{GLAM} professionals
        \item Computer and information scientists
        \item Funders
        \item Institutional management and leadership 
    \end{itemize}
\end{multicols}

Members of these stakeholder groups are working on software citation, in multiple overlapping directions resulting in competing efforts. 

For instance, is \enquote{software citation} about the practice of acknowledging the use of software, or is it about standardizing the ways in which citation is enabled? Along similar lines, is citation uptake driven by the need to acknowledge computational methods, regardless of where they were derived from, or is citation motivated by the need to ensure that people get credit? These perspectives imply different and incompatible approaches. 

Establishing a common vision regarding software citation will be a challenging, but critical undertaking because it will help all of the stakeholder communities make progress on software citation in a single agreed-upon direction.

To this end, Group 1 found that it will be necessary to develop a declaration that conforms to the original Software Citation Principles \cite{smith_sw_cit_prin_2016} and is broad enough to foster impact across all disciplines, while at the same time being more specific in how the principles should be implemented. The vision will specifically need to dictate two aspects of future work: a) what the researchers need to do and b) what research infrastructure needs to support. 

The vision should be aligned with other movements (e.g., FAIR and FAIR4RS practices, RSE community, Open Science, FAIR Digital Objects or FDOs) and include mechanisms for getting broad feedback on an initial declaration from different stakeholder communities – this will be important for getting consensus on the goals and benefits of software citation.

With these things in mind, the group agreed that the following goals should all be part of the conversation about a shared vision for the future of software citation:

\begin{multicols}{2}
\begin{itemize}
    \item Software should be shared long before, or independently of, other publications.
    \item Shared software should be persistent, citable, and FAIR \cite{chue_hong_fair4rs_2022}.
    \item Software should be shared in repositories.
    \item Scholarly credit should be given for the development and release of software.
    \item Software should be curated.
\end{itemize}
\end{multicols}

The group also identified the following more specific implementation issues a vision should address about software:

\begin{multicols}{2}
\begin{itemize}
    \item We need to decrease barriers to software publication and make clear what it means to publish.
    \item Needs to be recognized as a valid grant output.
    \item Not all sw. needs to be shared or published.
    \item Commercial and/or proprietary software should also be citable.
    \item It is extremely rare for software to be curated by specialists prior to publication.
\end{itemize}
\end{multicols}

The group agreed that the vision itself would eventually need to be laid out in a paper or an editorial. The question of \enquote{who takes this work on} was somewhat open but the group agreed that the \href{https://force11.org/groups/software-citation-implementation-working-group}{FORCE11 Software Citation Implementation Working Group} would be a good place to begin this initiative.

%\clearpage
\subsection{Group 2: Actions that can be taken without additional resources}\label{subsec:outcome:group2}
The second group of participants spent their time talking through actions seen as most impactful without access to significantly more resources (e.g., time, money, or buy-in from stakeholders). The three ideas below were identified as the most actionable and relatively short-term priorities for the group to pursue:

\begin{enumerate}
    \item Create a website that can act as a central source of information on software citation \textsl{in practice}.
    \begin{itemize}
        \item There are multiple places online (\cite{chivvis_citesoftwareorg_2022,druskat_cite_researchsw_2018}), where more \enquote{theoretical} work on software citation is shared (i.e., work done by workshop attendees) but there is no one place where different stakeholder communities can find straightforward instructions about software citation. 
        \item Members of Group 2 had already begun working on websites that somewhat meet this need, so the group decided to consolidate these efforts and establish a plan for the development, maintenance, and governance of a single website: \url{http://citesoftware.org}
    \end{itemize}
    
    \item Share a very simple recommendation on software citation file formats: software creators should create CFF files, and all other stakeholders should create CodeMeta files.
    \begin{itemize}
        \item This recommendation is based on the idea that CFF files are simple to create and maintain, focus solely on citation-relevant metadata and can be used to populate the more comprehensive CodeMeta files. While CodeMeta files as providers of descriptive metadata with a scope beyond software citation are extremely useful for enabling software preservation and reuse, they are not specifically intended to enable software citation. CFF files are also now recognized by GitHub and Zenodo/InvenioRDM (\href{https://gitlab.com/gitlab-org/gitlab/-/issues/337368 }{GitLab has an open request}), which will likely make their adoption more ubiquitous over time.
    \end{itemize}

    \item Write a paper for the \gls{GLAM} community on software-ready repositories and how they can enable software citation.
    \begin{itemize}
        \item Although this idea has a relatively long timescale, some members of the group had already begun engaging on this topic and it was clear that no additional resources were needed for this work to continue.
    \end{itemize}
\end{enumerate}

Although the group chose to focus on these three specific projects, the following ideas were also seen as worthwhile and in need of sharing:

\begin{multicols}{2}
\begin{itemize}
    \item Create guidance for software creators to help them have conversations about who to include as authors in their metadata.
    \item Create messaging materials for institutional policymakers to encourage software citation and the recognition of software as a scholarly output.
    \item Create short modules on software citation that could be incorporated into bibliographic instruction sessions in academic libraries.
    \item Write a letter that could be circulated to various funders advocating for actionable software preservation and citation policies, as well as funding opportunities specifically geared toward software development and maintenance.
    \item Develop a survey of repositories to help us determine the availability of software citation metadata (e.g., software version and release date fields) and the challenges repository managers face in implementing changes.
    \item Develop a survey of publishers to help us better understand the challenges that editors face in implementing software citation policies.
\end{itemize}
\end{multicols}

%\clearpage
\subsection{Group 3: Interdisciplinary Literature Guidance Outcomes}\label{subsec:outcome:group3}
Group 3 started by brainstorming potential solutions for mixed data deposits into (data) repositories. The group defined mixed data deposits as any deposit that contains research software code and/or accompanying data results or other non-software code materials.\footnote{This was a colloquially agreed upon definition and not one that was pulled from a specific outside source. The group is aware that we need engagement with a broader community to reach a definition covering the different perspectives and understandings (e.g., RO-Crate; RDA CURE-FAIR working group; Turing Way).}

Several ideas were presented, including: a) the need for single points of deposit and discovery, b) the ability to cite the entire deposit and separate into parts, and c) the need to identify each type of research software using best practices. Most of the ideas proved hard to implement and would require buy-in from data repository vendors to change their software. It was not feasible to draft a plan of action for many of these ideas.

Throughout this exercise, the group found themselves defining concepts and terminology as used in their field. The group consisted of members who were immersed in various fields such as research science, scholarly communications, and \gls{GLAM} fields. Due to these barriers of understanding, one suggestion was to create new resources for various stakeholder groups using language that the addressed group would easily comprehend. This would allow diverse stakeholders such as \gls{GLAM} professionals, data scientists across disciplines, and people in industry to exchange information and learn about software citation more seamlessly.

The group therefore pivoted from solutions for depositing mixed data types, to the creation of short-form documentation relating to subjects like software citation metadata, software publishing, and FAIR and FAIR4RS principles using language that groups such as archivists, students, and research software creators and users would understand.

Further ideas by the group outlined need for a software citation-related glossary and an online decision tree that would direct stakeholders to different resources depending on their selected audience and specific topic of interest. 

Group 3 eventually decided that in order to move forward they would need to establish an initial organizing committee that would reach out to other relevant organizations and undertake a literature review to inform the language used in resources for different stakeholders groups. These resources would subsequently be incorporated into the unified web presence proposed by Group 2, see \ref{subsec:outcome:group2}.

\section{Next Steps}
Workshop participants regrouped after the brainstorming sessions to share the outcomes of the structured exercises. Common themes emerged during the debrief including the need to house further work under a larger organizational umbrella, and the inclusion of stakeholders from different fields to improve software citation implementation.

To this end, participants agreed that the \href{https://force11.org/groups/software-citation-implementation-working-group}{FORCE11 Software Citation Implementation Working Group (SCIWG)} would be an appropriate convening entity to oversee the execution of future activities and make appropriate connections with other groups. Therefore, the workshop organizers agreed to bring the following actions to the SCIWG and to propose that work toward them be prioritized for the coming year: 

\begin{enumerate}
    \item Establish a unified vision for software citation
    \item Develop a single web presence to house resources related to software citation
    \item Promote simple guidance on the use of CFF and CodeMeta files
\end{enumerate}

\subsection{Work in Progress}
Workshop participants from Group 2 (see \ref{subsec:outcome:group2}) have already begun working to consolidate content from multiple websites. They are also prototyping a \enquote{decision tree} (or a flow chart) to help software creators make their software citable and to give guidance to people who want to cite software. This resource will inform the implementation and content discovery on the new website at \url{http://citesoftware.org}.

Participants from Group 3 (see \ref{subsec:outcome:group3}) also began developing ideas for a pathway to help different stakeholders identify the resources that would be most helpful to them.

\clearpage
\appendix

\addsec{Figures}\label{tables-figures}

\addsec{Acknowledgments}\label{acknowledgments}

We would like to thank \href{https://osf.io/hxy5k}{all of our participants} for their insight and for making time for our online and in-person events. We are also grateful to our facilitator and graphic recorder, who made our online sessions as productive as possible. Thank you too to the Institute for Museum and Library Services for funding this work, and Daniel Chivvis for leading the organization of events.

\setglossarystyle{altlist}
\printglossaries

\phantomsection
\printbibliography[heading=bibintoc]

@article{bouquin_credit_2020,
	title = {Credit {Lost}: {Two} {Decades} of {Software} {Citation} in {Astronomy}},
	volume = {249},
	issn = {0067-0049},
	shorttitle = {Credit {Lost}},
	doi = {10.3847/1538-4365/ab7be6},
	number = {1},
	journal = {The Astrophysical Journal Supplement Series},
	author = {Bouquin, Daina R. and Chivvis, Daniel A. and Henneken, Edwin and Lockhart, Kelly and Muench, August and Koch, Jennifer},
	month = jun,
	year = {2020},
	note = {Publisher: The American Astronomical Society},
	pages = {8},
}

@article{howison_sw_lit_2016,
	title = {Software in the scientific literature: {Problems} with seeing, finding, and using software mentioned in the biology literature},
	volume = {67},
	issn = {2330-1643},
	shorttitle = {Software in the scientific literature},
	doi = {10.1002/asi.23538},
	number = {9},
	journal = {Journal of the Association for Information Science and Technology},
	author = {Howison, James and Bullard, Julia},
	year = {2016},
	pages = {2137--2155},
}

@misc{chue_hong_ssi_journals_2012,
	title = {In which journals should {I} publish my software?},
	url = {https://www.software.ac.uk/which-journals-should-i-publish-my-software},
	journal = {In which journals should I publish my software?},
	author = {Chue Hong, Neil P.},
	year = {2012},
}

@article{van_noorden_top100_2014,
	title = {The top 100 papers},
	volume = {514},
	doi = {10.1038/514550a},
	language = {en},
	number = {7524},
	journal = {Nature News},
	author = {Van Noorden, Richard and Maher, Brendan and Nuzzo, Regina},
	month = oct,
	year = {2014},
	pages = {550},
}

@article{smith_joss_2018,
	title = {Journal of {Open} {Source} {Software} ({JOSS}): design and first-year review},
	volume = {4},
	issn = {2376-5992},
	shorttitle = {Journal of {Open} {Source} {Software} ({JOSS})},
	doi = {10.7717/peerj-cs.147},
	journal = {PeerJ Computer Science},
	author = {Smith, Arfon M. and Niemeyer, Kyle E. and Katz, Daniel S. and Barba, Lorena A. and Githinji, George and Gymrek, Melissa and Huff, Kathryn D. and Madan, Christopher R. and Mayes, Abigail Cabunoc and Moerman, Kevin M. and Prins, Pjotr and Ram, Karthik and Rokem, Ariel and Teal, Tracy K. and Guimera, Roman Valls and Vanderplas, Jacob T.},
	month = feb,
	year = {2018},
	note = {Publisher: PeerJ Inc.},
	pages = {e147},
}

@article{smith_sw_cit_prin_2016,
	title = {Software citation principles},
	volume = {2},
	issn = {2376-5992},
	doi = {10.7717/peerj-cs.86},
	number = {e86},
	journal = {PeerJ Computer Science},
	author = {Smith, Arfon M. and Katz, Daniel S. and Niemeyer, Kyle E. and {FORCE11 Software Citation Working Group}},
	year = {2016},
}

@misc{katz_f11_sciwg_2022,
	title = {{FORCE11} {Group}: {Software} {Citation} {Implementation} {Working} {Group}},
	url = {https://force11.org/groups/software-citation-implementation-working-group/},
	journal = {Software Citation Implementation Working Group},
	author = {Katz, Daniel S. and Fenner, Martin and Chue Hong, Neil P.},
	month = oct,
	year = {2022},
}

@misc{katz_sw_impl_chall_2019,
	title = {Software {Citation} {Implementation} {Challenges}},
	doi = {10.48550/arXiv.1905.08674},
	publisher = {arXiv},
	author = {Katz, Daniel S. and Bouquin, Daina and Hong, Neil P. Chue and Hausman, Jessica and Jones, Catherine and Chivvis, Daniel and Clark, Tim and Crosas, Mercè and Druskat, Stephan and Fenner, Martin and Gillespie, Tom and Gonzalez-Beltran, Alejandra and Gruenpeter, Morane and Habermann, Ted and Haines, Robert and Harrison, Melissa and Henneken, Edwin and Hwang, Lorraine and Jones, Matthew B. and Kelly, Alastair A. and Kennedy, David N. and Leinweber, Katrin and Rios, Fernando and Robinson, Carly B. and Todorov, Ilian and Wu, Mingfang and Zhang, Qian},
	month = may,
	year = {2019},
	note = {arXiv:1905.08674 [cs]},
}

@article{wilkinson_fair_2016,
	title = {The {FAIR} {Guiding} {Principles} for scientific data management and stewardship},
	volume = {3},
	copyright = {2016 The Author(s)},
	issn = {2052-4463},
	doi = {10.1038/sdata.2016.18},
	number = {1},
	journal = {Scientific Data},
	author = {Wilkinson, Mark D. and Dumontier, Michel and Aalbersberg, IJsbrand Jan and Appleton, Gabrielle and Axton, Myles and Baak, Arie and Blomberg, Niklas and Boiten, Jan-Willem and da Silva Santos, Luiz Bonino and Bourne, Philip E. and Bouwman, Jildau and Brookes, Anthony J. and Clark, Tim and Crosas, Mercè and Dillo, Ingrid and Dumon, Olivier and Edmunds, Scott and Evelo, Chris T. and Finkers, Richard and Gonzalez-Beltran, Alejandra and Gray, Alasdair J. G. and Groth, Paul and Goble, Carole and Grethe, Jeffrey S. and Heringa, Jaap and ’t Hoen, Peter A. C. and Hooft, Rob and Kuhn, Tobias and Kok, Ruben and Kok, Joost and Lusher, Scott J. and Martone, Maryann E. and Mons, Albert and Packer, Abel L. and Persson, Bengt and Rocca-Serra, Philippe and Roos, Marco and van Schaik, Rene and Sansone, Susanna-Assunta and Schultes, Erik and Sengstag, Thierry and Slater, Ted and Strawn, George and Swertz, Morris A. and Thompson, Mark and van der Lei, Johan and van Mulligen, Erik and Velterop, Jan and Waagmeester, Andra and Wittenburg, Peter and Wolstencroft, Katherine and Zhao, Jun and Mons, Barend},
	month = mar,
	year = {2016},
	pages = {160018},
}

@article{chue_hong_fair4rs_2022,
	title = {{FAIR} {Principles} for {Research} {Software} ({FAIR4RS} {Principles})},
	doi = {10.15497/RDA00068},
	author = {Chue Hong, Neil P. and Katz, Daniel S. and Barker, Michelle and Lamprecht, Anna-Lena and Martinez, Carlos and Psomopoulos, Fotis E. and Harrow, Jen and Castro, Leyla Jael and Gruenpeter, Morane and Martinez, Paula Andrea and Honeyman, Tom and Struck, Alexander and Lee, Allen and Loewe, Axel and van Werkhoven, Ben and Jones, Catherine and Garijo, Daniel and Plomp, Esther and Genova, Francoise and Shanahan, Hugh and Leng, Joanna and Hellström, Maggie and Sandström, Malin and Sinha, Manodeep and Kuzak, Mateusz and Herterich, Patricia and Zhang, Qian and Islam, Sharif and Sansone, Susanna-Assunta and Pollard, Tom and Atmojo, Udayanto Dwi and Williams, Alan and Czerniak, Andreas and Niehues, Anna and Fouilloux, Anne Claire and Desinghu, Bala and Goble, Carole and Richard, Céline and Gray, Charles and Erdmann, Chris and Nüst, Daniel and Tartarini, Daniele and Ranguelova, Elena and Anzt, Hartwig and Todorov, Ilian and McNally, James and Moldon, Javier and Burnett, Jessica and Garrido-Sánchez, Julián and Belhajjame, Khalid and Sesink, Laurents and Hwang, Lorraine and Tovani-Palone, Marcos Roberto and Wilkinson, Mark D. and Servillat, Mathieu and Liffers, Matthias and Fox, Merc and Miljković, Nadica and Lynch, Nick and Martinez Lavanchy, Paula and Gesing, Sandra and Stevens, Sarah and Martinez Cuesta, Sergio and Peroni, Silvio and Soiland-Reyes, Stian and Bakker, Tom and Rabemanantsoa, Tovo and Sochat, Vanessa and Yehudi, Yo and WG, RDA FAIR4RS},
	month = may,
	year = {2022},
}

@article{katz_fresh_look_2021,
	title = {A {Fresh} {Look} at {FAIR} for {Research} {Software}},
	journal = {arXiv:2101.10883 [cs]},
	author = {Katz, Daniel S. and Gruenpeter, Morane and Honeyman, Tom and Hwang, Lorraine and Wilkinson, Mark D. and Sochat, Vanessa and Anzt, Hartwig and Goble, Carole and 1, for FAIR4RS Subgroup},
	month = feb,
	year = {2021},
	note = {arXiv: 2101.10883},
}

@article{malone_softonto_2014,
  title        = {The Software Ontology (SWO): a resource for reproducibility in biomedical data analysis, curation and digital preservation},
  author       = {Malone, James and Brown, Andy and Lister, Allyson L. and Ison, Jon and Hull, Duncan and Parkinson, Helen and Stevens, Robert},
  year         = 2014,
  month        = {6},
  journal      = {Journal of Biomedical Semantics},
  volume       = 5,
  number       = 1,
  pages        = 25,
  doi          = {10.1186/2041-1480-5-25},
  url          = {https://doi.org/10.1186/2041-1480-5-25},
  abstractnote = {Biomedical ontologists to date have concentrated on ontological descriptions of biomedical entities such as gene products and their attributes, phenotypes and so on. Recently, effort has diversified to descriptions of the laboratory investigations by which these entities were produced. However, much biological insight is gained from the analysis of the data produced from these investigations, and there is a lack of adequate descriptions of the wide range of software that are central to bioinformatics. We need to describe how data are analyzed for discovery, audit trails, provenance and reproducibility.}
}

@misc{garjo_softdesconto_2021,
  title        = {The Software Description Ontology},
  author       = {Garijo, Daniel and Ratnakar, Varun and Gil, Yolanda and Khider, Deborah},
  year         = 2021,
  month        = {5},
  journal      = {The Software Description Ontology},
  url          = {https://w3id.org/okn/o/sd},
  abstractnote = {An ontology for describing software components, including their metadata (attribution, licensing, usage instructions, how to get support) and their inputs, outputs and variables. The ontology extends schema.org and CodeMeta vocabularies and is based on OntoSoft, which proposed a vocabulary for describing software by asking scientists questions.}
}

@book{wilder_doap_2021,
  title        = {DOAP: Description Of A Project},
  author       = {Wilder-James, Edd},
  year         = 2021,
  month        = {11},
  url          = {https://github.com/ewilderj/doap},
  abstractnote = {RDF schema for describing software projects}
}

@article{wuersch_seon_2012,
  title        = {SEON: a pyramid of ontologies for software evolution and its applications},
  author       = {Würsch, Michael and Ghezzi, Giacomo and Hert, Matthias and Reif, Gerald and Gall, Harald C.},
  year         = 2012,
  month        = {11},
  journal      = {Computing},
  volume       = 94,
  number       = 11,
  pages        = {857–885},
  doi          = {10.1007/s00607-012-0204-1},
  abstractnote = {The Semantic Web provides a standardized, well-established frameworkto define and work with ontologies. It is especially apt for machineprocessing. However, researchers in the field of software evolutionhave not really taken advantage of that so far. In this paper, weaddress the potential of representing software evolution knowledgewith ontologies and Semantic Web technology, such as Linked Data andautomated reasoning. We present Seon, a pyramid of ontologies forsoftware evolution, which describes stakeholders, their activities,artifacts they create, and the relations among all of them. We showthe use of evolution-specific ontologies for establishing a sharedtaxonomy of software analysis services, for defining extensiblemeta-models, for explicitly describing relationships amongartifacts, and for linking data such as code structures, issues(change requests), bugs, and basically any changes made to a systemover time. For validation, we discuss three different approaches,which are backed by Seon and enable semantically enriched softwareevolution analysis. These techniques have been fully implemented astools and cover software analysis with web services, a naturallanguage query interface for developers, and large-scale softwarevisualization.}
}

@book{jones_codemeta_2017,
  title        = {CodeMeta: an exchange schema for software metadata. Version 2.0},
  author       = {Jones, Matthew B. and Boettiger, Carl and Mayes, Abby Cabunoc and Smith, Arfon and Slaughter, Peter and Niemeyer, Kyle and Gil, Yolanda and Fenner, Martin and Nowak, Krzysztof and Hahnel, Mark and Coy, Luke and Allen, Alice and Crosas, Mercè and Sands, Ashley and Hong, Neil Chue and Cruse, Patricia and Katz, Dan and Goble, Carole},
  year         = 2017,
  doi          = {10.5063/schema/codemeta-2.0},
  note         = {Published: KNB Data Repository}
}

@article{druskat_cff_2021,
  title        = {Citation File Format},
  author       = {Druskat, Stephan and Spaaks, Jurriaan H. and Chue Hong, Neil and Haines, Robert and Baker, James and Bliven, Spencer and Willighagen, Egon and Pérez-Suárez, David and Konovalov, Alexander},
  year         = 2021,
  month        = {8},
  doi          = {10.5281/ZENODO.5171937},
  note         = {Publisher: Zenodo Version Number: 1.2.0},
  abstractnote = {CITATION.cff files are plain text files with human- and machine-readable citation information for software. Code developers can include them in their repositories to let others know how to correctly cite their software. This is the specification for the Citation File Format.}
}

@misc{chivvis_citesoftwareorg_2022,
	title = {CfA-Library/citesoftware.org: Your go-to guide for software citation.},
	url = {https://github.com/CfA-Library/citesoftware.org},
	urldate = {2023-02-05},
	author = {Chivvis, Daniel},
	month = aug,
	year = {2022},
}

@misc{druskat_cite_researchsw_2018,
	title = {Research {Software} {Citation}: {Cite} and {Make} {Citable}},
	url = {https://cite.research-software.org/},
	urldate = {2023-02-05},
	journal = {Research Software Citation},
	author = {Druskat, Stephan and Crusoe, Michael R.},
	month = feb,
	year = {2018},
}

@misc{hermes_2022,
	title = {Software publications with rich metadata: state of the art, automated workflows and {HERMES} concept},
	shorttitle = {Software publications with rich metadata},
	url = {http://arxiv.org/abs/2201.09015},
	doi = {10.48550/arXiv.2201.09015},
	publisher = {arXiv},
	author = {Druskat, Stephan and Bertuch, Oliver and Juckeland, Guido and Knodel, Oliver and Schlauch, Tobias},
	month = jan,
	year = {2022},
	note = {arXiv:2201.09015 [cs]},
}

\end{document}